\documentclass[preprint,floats,aps,epsfig,nofootinbib,amssymb]{JHEP3}
\usepackage{graphicx}
\usepackage{bm}
\usepackage{amsmath}
\usepackage{amssymb}
\usepackage{amsthm}
\newcommand{\aop}{{ \mathbf{a}}}
\newcommand{\bop}{{\mathbf b}}

\newcommand{\de}{\delta}

\newcommand{\La}{\Lambda}

\newcommand{\om}{\omega}

\newcommand{\Si}{\Sigma}

\newcommand{\ra}{\rightarrow}

\newcommand{\be}{\begin{equation}}
\newcommand{\ee}{\end{equation}}
\newcommand{\bea}{\begin{eqnarray}}
\newcommand{\eea}{\end{eqnarray}}
\newcommand{\bean}{\begin{eqnarray*}}
\newcommand{\eean}{\end{eqnarray*}}
\newcommand{\dd}{\partial}

\newcommand{\LCDM}{$\Lambda$\rm{CDM}}

\newcommand{\diff}{{\rm d}}

\newcommand{\lsim}   {\mathrel{\mathop{\kern 0pt \rlap
  {\raise.2ex\hbox{$<$}}}
  \lower.9ex\hbox{\kern-.190em $\sim$}}}
\newcommand{\gsim}   {\mathrel{\mathop{\kern 0pt \rlap
  {\raise.2ex\hbox{$>$}}}
  \lower.9ex\hbox{\kern-.190em $\sim$}}}

\title{A longitudinal gauge degree of freedom and the Pais Uhlenbeck field}

\author{{Jose} Beltr\'an Jim\'enez$^{a,b}$, {Enea} Di Dio$^{b}$ and {Ruth} Durrer$^{b}$\\
${^a}${Centre for Cosmology, Particle Physics and Phenomenology,\\
Institute of Mathematics and Physics, Louvain University,\\ 2 Chemin du Cyclotron, 1348~Louvain-la-Neuve~(Belgium)}\\
\\
$^b${D\'epartement de Physique Th\'eorique and Center for Astroparticle Physics,\\
Universit\'e de Gen\`eve, 24 quai Ansermet, CH--1211 Gen\`eve 4,
Switzerland.}
\\
\\
\email{jose.beltran@uclouvain.be}, \quad \email{enea.didio@unige.ch}, \quad
\email{ruth.durrer@unige.ch}
}

\date{\today}

\abstract{
We show that a longitudinal gauge degree of freedom for a vector field is equivalent to a Pais-Uhlenbeck scalar field. With the help of this equivalence, we can determine natural interactions of this field with scalars and fermions. Since the theory has a global $U(1)$ symmetry, we have the usual conserved current of the charged fields, thanks to which the dynamics of the scalar field is not modified by the interactions. We use this fact to  consistently quantize the theory even in the presence of interactions. We argue that such a degree of freedom can only be excited by gravitational effects like the inflationary era of the early universe and may play the role of dark energy in the form of an effective cosmological constant whose value is linked to the inflation scale.
}

\keywords{Cosmology: theory, dark energy, higher derivative theories, quantization}

\begin{document}

\section{Introduction}\label{s:intro}
The standard model of cosmology  provides a very successful description of our Universe~\cite{Komatsu}. However, it is based on the idea that about 96\%
of the energy density in the Universe stems from particles and fields which are not part of the standard model of particle physics and have never been observed in colliders. Dark matter, which amounts to about 26\% of the Universe content, at least has the properties of typical non-relativistic particles, but
dark energy, the component which makes up about 70\% of the energy density of the Universe must be endowed with very exotic properties, like a strong negative pressure, in order to explain the observed current acceleration of the
expansion of the Universe. This most unexpected discovery has been awarded the 2011 Nobel prize in 
physics~\cite{nobel11}. 

Even if a cosmological constant has the right properties and is in agreement with cosmological observations, 
 this solution is unsatisfactory from a purely theoretical point of view, as its associated scale
$(\rho_\La)^{1/4} \simeq 10^{-3}{\rm eV}$ is so much smaller than the natural scale of gravity given by $M_p\simeq 10^{18}$ GeV. On the other hand, one could expect this scale to be related to some cutoff
scale coming from particle physics, which should be, at least, the scale of supersymmetry $E_{\rm susy}> 1$TeV, again much larger than the value of the cosmological constant  inferred from observations (see \cite{Martin:2012bt} for an extensive discussion about the cosmological constant problem).

This fine tuning issue of the cosmological constant has led the community to search for different solutions to the problem of accelerated expansion. 
Different dark energy models like e.g. quintessence or large scale modifications of gravity have been 
explored~\cite{deRR}.

Researchers have also looked into theories with higher derivatives in the Lagrangian density, like e.g.  the so-called $f(R)$ theories \cite{f(R)} or the galileon 
field~\cite{galileon}. Theories with higher derivatives look dangerous at first sight because they typically lead to the presence of ghosts (particles with negative kinetic energy) reflecting the Ostrogradski instability~\cite{Ostro,Woody}. This instability appears because these theories lead to higher than second order equations of motion, so that new degrees of freedom appear and they are usually ghost-like. It has been shown 
recently~\cite{Chen:2012au}, that trying to eliminate the ghost by introducing additional constraints  does not work in general, unless the corresponding phase space gets dimensionally-reduced.  Even though without the presence of any interactions, 
the energy of a given field remains constant and the instability cannot develop, we expect that the coupling of such a 
field to other degrees of freedom or quantum effects will potentially provoke uncurable instabilities.

However, exceptions to this generic rule exist in degenerate theories like the aforementioned $f(R)$ theories of gravity and, more recently, the galileon fields where, thanks to the specific structure of the interactions, the equations remain of second order even though higher derivative terms are present in the action. 

Interestingly, 
the dark energy problem, forces us to re-think our boundaries of what we can accept as  sensible (effective) 
physical theories~\cite{ArkaniHamed}. This is the youngest example of the cross fertilization of cosmological observations and fundamental theoretical physics.

The present paper inscribes in this framework. Here we show that the action for a vector field having only a residual gauge symmetry, which has already been proposed as a candidate for dark energy \cite{Jimenez:2009dt}, can be rewritten as an ordinary $U(1)$ gauge vector field theory together with a 
 (degenerate) Pais-Uhlenbeck (PU) scalar field, the field version of the PU oscillator first discussed in ~\cite{Pais:1950za}.  Even though this is a higher derivative oscillator, it has been 
shown in~\cite{Bender:2007wu} that this model can be quantized with a positive spectrum for the Hamiltonian because it corresponds to a special class of Hamiltonians that, although being non-hermitian, exhibit a ${\mathcal PT}$ symmetry that allows to construct a  quantum theory without negative energy states or a unitary evolution.  However, such a construction relies on the presence of non-hermitian operators so that the classical limit of the theory remains unclear. In fact, the classical Hamiltonian is still unbounded and a prescription to go from the quantum theory to the classical solutions is lacking. Moreover, although it is an interesting construction to have a bounded spectrum for the quantum theory, the problem with higher derivative terms actually becomes manifest when interactions are introduced \cite{Smilga:2008pr}. As a matter of fact,  the free theory is not sick and provides a unitary evolution, even though the Ostrogradski ghostly degree of freedom is present. For special types of interactions, unitarity might be maintained  \cite{Smilga:2008pr,Smilga:2004cy}.

Here we derive an alternative way to consistently quantize the degenerate PU field  that makes use of the presence of another symmetry in which the field can be shifted by an arbitrary harmonic function. We use this symmetry to restrict the physical Hilbert space or to fix the gauge and quantize only the healthy physical mode. Then, we include interactions with charged scalars or fermions, in a way which is motivated by the 
interpretation of this field, and we show that our restriction of the physical states remains intact.

The remainder of this paper is organized  as follows: In the next section we show the relation of the vector field action with the residual gauge symmetry and 
the PU field. In Section~\ref{s:inter} we include interactions. In Section~\ref{s:disc} we discuss the quantization of the theory and show how it could play the role of dark energy or of the inflaton. (Even if in its present form the model does not propose a mechanism to end  inflaton.) 
In Section~\ref{s:con} we conclude and
discuss further investigations which can be performed to compare this model with standard \LCDM.

\section{The St\"uckelberg trick and the Pais Uhlenbeck field}\label{s:StPU}

\subsection{From the Pais Uhlenbeck oscillator to the Pais Uhlenbeck field}

We shall start by briefly introducing the PU oscillator~\cite{Pais:1950za} and showing how it can arise as the Fourier modes of a certain higher order field theory. The PU oscillator is described by the following action:
\be \label{PUaction}
S_{PU}=\frac\gamma 2\int \diff t\left[\ddot{z}^2-(\omega_1^2+\omega_2^2)\dot{z}^2+\omega_1^2\omega_2^2z^2\right],
\ee
where $\gamma$ is an arbitrary parameter. Since the action depends on the second time derivative, this model is expected to exhibit the Ostrogradski instability. The corresponding Hamiltonian can be derived either by using the Dirac method~\cite{Dirac} or directly from the Ostrogradski Hamiltonian~\cite{Ostro} as defined for higher order derivative theories. In either case, one obtains the expression
\be\label{HPU}
H_{PU}\left( z, x, p_z, p_x \right)=\frac{p_x^2}{2\gamma}+p_z x+\frac{\gamma}{2}(\omega_1^2+\omega_2^2)x^2-\frac \gamma2 \omega_1^2\omega_2^2z^2,
\ee
where $p_z$ and $p_x$ are the canonical conjugate momenta of the canonical variables $z$ and $x\equiv\dot{z}$ respectively. The term linear in the conjugate momentum,  $p_zx $, represents the previously advertised Ostrogradski instability.  At the classical level the instability may appear only by coupling the PU oscillator to other systems. The action~(\ref{PUaction}) leads to the  fourth order equation of motion
\be \label{PUeom}
\frac{\diff^4z}{\diff t^4}+(\omega_1^2+\omega_2^2)\frac{\diff^2z}{\diff t^2}+\omega_1^2\omega_2^2z=0,
\ee
which is solved by the superposition of two modes with frequencies $\omega_1$ and $\omega_2$ in the non-degenerate case ($\omega_1\neq \omega_2$):
\be
z(t)=a_1e^{-i\omega_1 t}+ a^*_1e^{i\omega_1 t}+a_2e^{-i\omega_2 t}+ a^*_2e^{i\omega_2 t}.
\ee
From this solution we see that no instabilities in the form of exponentially growing modes appear in the classical solutions even though the Hamiltonian has the aforementioned linear term associated with the Ostrogradski instability. This should not be surprising, since, at the classical level, only the introduction of interactions can develop the Ostrogradski instability by shifting the poles of the corresponding propagator off the real axis. Note the difference of these oscillating solutions to
tachyons (i.e., modes with negative $\omega^2$) where the above solutions would grow exponentially. 

On the other hand, in the degenerate case ($\omega_1= \omega_2$) we find
\be
z\left( t \right) = \left( c_1 + c_2 t \right) e^{-i \omega t} +\left(  c^*_1 +  c^*_2 t \right) e^{i \omega t},\label{degsol}
\ee
where the term linear in $t$ comes from the fact that the roots of the characteristic equation for the equation of motion~(\ref{PUeom}) are degenerate (or, in other words, the propagator has a double pole). Even though this term is growing in time, its growth is milder than an exponentially growing mode. In particular there is no imaginary propagation speed that would signify  a classical instability in the form of a tachyonic mode. This linear growth is independent of the presence of the Ostrograski instability, that is still present in the theory and can develop when  interactions are introduced. 

In summary, we have seen that the PU oscillator does not have tachyonic instabilities, although the Ostrogradski instability is still present and could turn the model out of control when interactions are introduced. 

In the literature the PU oscillator has been already widely studied (cf.~\cite{Pais:1950za,Bender:2007wu,Smilga:2008pr, PUoscillator}). Here we are interested in the field version of the PU model. Let us consider the action
\begin{eqnarray}\label{PUactionfield}
S&=&\xi \int \diff^4x \Big[ \phi\left( \Box +m_1^2 \right) \left( \Box  + m_2^2 \right) \phi \Big]\nonumber\\
&=&\xi\int\diff^4x\Big[(\Box\phi)^2-(m_1^2+m_2^2)\partial_\mu\phi\partial^\mu\phi+m_1^2m_2^2\phi^2\Big]
\end{eqnarray}
where $\Box =\partial_\mu\partial^\mu$ is the d'Alembertian operator and $\xi$ a dimensionless parameter whose value can be fixed by the normalization of $\phi$.  Note that the PU 
scalar field $\phi$ defined here is dimensionless in four spacetime dimensions.

We now show that each spatial Fourier mode of the PU field describes a PU oscillator where the frequencies $\omega_1$ and $\omega_2$ are determined by the masses $m_1$ and $m_2$. From the action~(\ref{PUactionfield}) we derive the equation of motion
\be
(\Box+m_1^2) \left( \Box  +m_2^2 \right) \phi =0.
\ee
Expanding the PU field in spatial Fourier modes
\be
\phi = \int \frac{\diff^3k}{\left( 2 \pi \right)^{3/2}}   \phi_k(t) e^{i \vec{k}\cdot\vec{x} }
\ee
the equation of motion for each mode $\phi_k$ becomes
\be \label{PUfieldeom}
\frac{\diff^4 \phi_k}{\diff t^4} + \Big( 2 k^2 + m_1^2+m_2^2 \Big) \frac{\diff^2  \phi_k}{\diff t^2} + \Big[ k^4 + k^2\left(m_1^2+m_2^2\right) +m_1^2m_2^2\Big] \phi_k =0.
\ee
Comparing Eqs.~(\ref{PUeom}) and~(\ref{PUfieldeom}) we identify
\be
\omega_1^2 + \omega_2^2 = 2 k^2 +m_1^2+m_2^2 \quad \text{and} \quad \omega_1^2 \omega_2^2 = k^4 + k^2 (m_1^2+m_2^2)+m_1^2m_2^2,
\ee
so that, as expected, the frequencies are given by:
\begin{eqnarray}
\omega_1^2=k^2+m_1^2\nonumber\\
\omega_2^2=k^2+m_2^2.
\end{eqnarray}
The action (\ref{PUactionfield}) describes two massive modes with positive masses $m_1$ and $m_2$ (none of them represents a tachyonic degree of freedom).   However, one of them is actually a ghost, with the sign of $m_1^2-m_2^2$  determining which one of the two modes is the ghost. In the case where one of these parameters vanishes, one of the modes becomes massless. This corresponds to the case without the quadratic potential in the action, i.e., only the derivative terms remain despite the fact that  one of the modes is massive. Finally, when both masses vanish $m_1=m_2=0$, we have a pure fourth order theory and the solutions of the characteristic equation are degenerate, i.e., both modes satisfy the usual dispersion relation for a massless mode. This is the aforementioned degeneracy leading to the term linearly growing with time appearing in (\ref{degsol}). Each Fourier mode of a massless PU field is described by a degenerate PU oscillator and analogously for the massive case.

\subsection{The St\"uckelberg trick }
In the following we shall show how the degenerate PU field can be identified with a St\"uckelberg field. This field was first introduced in the Proca action for a massive vector field to restore the $U(1)$ gauge invariance of the theory~\cite{Stuck}. Thus, the vector field could acquire a non-vanishing mass and still preserve the gauge symmetry in a somewhat simplified version of the Higgs mechanism. Here, we shall start from an action for a massless vector field, but with a modified kinetic term,
\be
S=\int \diff^4x\left[-\frac{1}{4}F_{\mu\nu}F^{\mu\nu}+\frac{1}{2}\xi(\partial_\mu A^\mu)^2\right] \,,\label{gfaction}
\ee
  with $F_{\mu\nu}=\partial_\mu A_\nu -\partial_\nu A_\mu$ and $\xi$ and arbitrary dimensionless parameter (that can be fixed by the normalization of the longitudinal component of the vector field).  In this action,  $-F^2/4$ is the usual Maxwell term, whereas the second term is usually introduced in the quantization of the theory as a gauge-fixing term and partially breaks the gauge symmetry. However, here we shall not consider it as a purely gauge-fixing term, but  as a fully physical term. This action is no longer invariant under general $U(1)$ transformations of the vector field, but only under residual gauge transformations
$A_\mu\rightarrow A_\mu+\partial_\mu\theta$ which satisfy $\Box\theta=0$. 
In addition to the usual transverse modes associated with $F_{\mu\nu}$, this action has a third  degree of freedom which appears here
as longitudinal vector mode\footnote{Here we refer to a 4-longitudinal mode  such that $A^\mu k_\mu\neq 0$, as opposed   with the usual terminology used in theories with massive vector fields where the longitudinal mode refers to the mode parallel to the 3-momentum $\vec{k}$.} and that is associated with $\partial_\mu A^\mu$. Since such a mode gives a scalar physical quantity, we can alternatively interpret it as a scalar  mode. 

In the past~\cite{Jimenez:2009dt}, it has been shown that this additional mode gives a contribution to the cosmological energy-momentum tensor   that is proportional to the metric tensor  so that its equation of state is $p=-\rho$  and it has therefore been suggested that this could play the role of the dark energy as an effective cosmological constant. So far, no evidences for longitudinal photons have been found in colliders and electromagnetism is unbroken in the real Universe to very high precision. Furthermore, there is no experimental evidence for the existence of an additional $U(1)$ gauge field in nature.

In this work, we show that when restoring  $U(1)$ gauge invariance using the St\"uckelberg trick, this longitudinal mode becomes a  scalar degree of freedom.
Interestingly, it does not correspond to a normal scalar field but to the previously introduced (degenerate) Pais-Uhlenbeck  field~\cite{Pais:1950za}, which has second time derivatives in the action and fourth order equations of motion. The Ostrogradski ghost of the PU field will be related to the ghostly degree of freedom introduced by the gauge-fixing-like term in (\ref{gfaction}). In subsequent sections we shall see how to deal with such a ghost and consistently quantize the theory.
 
Let us now restore $U(1)$ gauge invariance by introducing a St\"uckelberg field. In other words, we replace $A_\mu\rightarrow A_\mu+\partial_\mu\phi$ so that the action becomes
\begin{eqnarray}
S&=&\int \diff^4x\left[-\frac{1}{4}F_{\mu\nu}F^{\mu\nu}+\frac{1}{2}\xi\left(\partial_\mu (A^\mu+\partial^\mu\phi)\right)^2\right]\nonumber\\
&=&\int \diff^4x\left[-\frac{1}{4}F_{\mu\nu}F^{\mu\nu}+\frac{1}{2}\xi\left(
(\partial_\mu A^\mu)^2+(\Box\phi)^2+2 \partial_\mu A^\mu\Box\phi
\right)\right] \,. \label{e:act2}
\end{eqnarray}
This action is fully gauge-invariant under gauge transformations which act on both, $A_\mu$ and $\phi$
\begin{eqnarray}
A_\mu&\rightarrow& A_\mu+\partial_\mu\Lambda,\nonumber\\
\phi&\rightarrow&\phi-\Lambda.
\end{eqnarray}
The action (\ref{gfaction}) can thus be interpreted as the action~({\ref{e:act2})  in a gauge such that  $\Box\phi=0$.  In such a gauge, the residual gauge symmetry of (\ref{gfaction}) remains because the condition $\Box\phi=0$ does not fix the gauge completely, but still is invariant under gauge transformations $A_\mu\rightarrow A_\mu+\partial_\mu\theta$, $\phi \ra \phi-\theta$ 
which satisfy $\Box\theta=0$.  This is analogous to the residual gauge symmetry that remains in standard electromagnetism after imposing the Lorenz gauge condition.   Since we have introduced the St\"uckelberg field to restore the full $U(1)$ gauge symmetry, one could also consider a mass term for the vector field, as in the original St\"uckelberg model. Then, one would obtain a massive PU field for the scalar field plus  additional couplings to the vector field. We will not explore this possibility here because we are only interested in studying the massless vector field case. A way to prohibit the mass term (that would be allowed by the required symmetries of our action) is to impose the additional symmetry for the St\"uckelberg field $\phi\rightarrow\phi+\vartheta$ with $\vartheta$ a harmonic function. Our action ({\ref{e:act2}) does fulfill this symmetry, whereas the term $A_\mu\partial^\mu\phi$ that is generated from the mass term does not.   

 The equations of motion obtained from action~(\ref{e:act2}) are
\begin{eqnarray}
\partial_\nu F^{\mu\nu}+\xi\partial^\mu\left[\partial_\nu A^\nu+ \Box\phi\right]=0 \,,\\
\Box\left(\Box\phi+\partial_\mu A^ \mu\right)=0 \,.
\end{eqnarray}
  Notice that the second equation is nothing but the 4-divergence of the first one.  If we now fix to Lorenz gauge $\partial_\mu A^\mu=0$, we are left with
\begin{eqnarray}
\Box A_\mu&=&\xi\partial_\mu\Box\phi \,,\\
\Box^2\phi&=&0 \,.
\end{eqnarray}
  The equation for the PU field completely decouples, whereas it acts as an effective external conserved current term for the vector field. Such an external source is determined by the gradient of $\Box\phi$ so that it only affects the longitudinal mode of $A_\mu$. In other words, the transverse modes completely decouple from $\phi$ and only the longitudinal mode is affected. Indeed, if we introduce the field $B_\mu\equiv A_\mu-\xi\partial_\mu\phi$, the equations can be written as
\begin{eqnarray}
\Box B_\mu&=&0  \,, \\
\Box^2\phi&=&0 \,
\end{eqnarray}
where now we have $\partial_\mu B^\mu=-\xi\Box\phi$ as the gauge condition. The PU field determines the longitudinal mode of the vector field $B_\mu$ which, in addition, satisfies a free wave equation. In fact, since we still have the residual gauge symmetry, we can use it to set $B_0=0$ so that we obtain $\nabla\cdot\vec{B}=\xi\Box\phi$.  In this gauge, we have the free wave equations of motion for the transverse modes of the vector field and for $\Box\phi$, whereas the longitudinal mode of the vector field is determined by the PU field (up to a residual gauge transformation). Notice that naively imposing the Lorenz gauge condition directly in the action (\ref{e:act2}) leads to
\begin{eqnarray}
S&=&\frac 12\int \diff^4x\left[- \partial_\mu A_\nu \partial^\mu A^\nu+\xi(\Box\phi)^2\right] \,.
\end{eqnarray}
This reproduces the correct equations of motion for the transversal modes of $A_\mu$ and for $\phi$, but it does not yield the correct equation for the longitudinal mode of the vector field. We can alternatively use a gauge such that\footnote{Since the quantity $\partial_\mu A^\mu+\Box\phi$ is gauge invariant, we cannot impose the condition $\partial_\mu A^\mu=-\Box\phi$. For our choice, this only happens in the limit $\xi\rightarrow\infty$.} $\partial_\mu A^\mu=-\frac{\xi}{1+\xi}\Box\phi$, which again leaves a residual gauge symmetry. Then, the equations of motion read:
\begin{eqnarray}
\Box A_\mu&=&0\,,\\
\Box^2\phi&=&0.
\end{eqnarray}
When imposing this gauge at the level of the action we obtain:
\be
S=\frac 12\int \diff^4x\left[- \partial_\mu A_\nu \partial^\mu A^\nu+\frac{\xi}{1+\xi}(\Box\phi)^2\right]
\ee
which reproduces the correct equations of motion and, if we additionally impose the used gauge as a subsidiary condition, we also obtain the correct relation between $\partial_\mu A^\mu$ and $\Box\phi$. In any case, it is more clear by looking at the equations of motion directly that we have completely decoupled transverse modes for the vector field satisfying free wave equations plus the degenerate PU field which determines the longitudinal mode of the vector field. It will be useful to note that neither of the two gauges discussed fixes  the gauge freedom completely, but they leave us with the residual gauge  symmetry $A_\mu\rightarrow A_\mu+\partial_\mu\theta$, $\phi \ra \phi-\theta$ with $\Box\theta=0$.  In addition to this residual symmetry, we also have the aforementioned symmetry that prevents the appearance of the mass term, i.e.  $A_\mu\rightarrow A_\mu$, $\phi \ra \phi+\vartheta$ with $\Box\vartheta=0$. In other words, we can perform a residual gauge transformation with two different (harmonic) gauge parameters $\theta$ and $\vartheta-\theta$ for $A_\mu$ and $\phi$ respectively. 

Interestingly, the transformation for the scalar field can be regarded as a generalized shift symmetry, since we can shift $\phi$ not only by a constant, like in the case of a Goldstone boson, or a linear function of the coordinates, like in the case of the galileon field, but by an arbitrary harmonic function. Of course, the constant shift and the Galilean transformations are particular cases of this more general symmetry.

We know that the Maxwell term describes a well behaved theory also at the quantum level and 
with interactions. However, the lack of stability under quantization is a common feature in higher order derivative Lagrangians. As we have already discussed, such theories suffer, for instance, from the Ostrogradski instability which implies that their Hamiltonian is not bounded from below~\cite{Ostro,Woody}.
It has been argued, however, that the Ostrogradski ghost instability which is present in the PU oscillator can be {\it cured} and that the theory can be consistently quantized thanks to the presence of an unbroken ${\mathcal PT}$ symmetry that allows non-hermitian Hamiltonians to lead to unitary evolution~\cite{Bender:2007wu}. However, how to take the classical limit of theories quantized within such a framework remains unclear. In Section~\ref{s:disc} we shall
come back to this issue and develop an alternative consistent quantization procedure for the PU action based on the  generalized shift symmetry and that makes no use of non-hermitian operators.

To end this Section, we want to mention that, in the same way that we can write the Maxwell Lagrangian in terms of the 2-form
${\mathcal F}=\diff{\mathcal A}= \frac12F_{\mu\nu}\diff x^\mu \wedge \diff x^\nu$, as $S_{\rm Maxwell}=-\frac 12\int {\mathcal F}\wedge *{\mathcal F} $ for ${\mathcal A} =A_\mu \diff x^\mu$, we find for the degenerate PU action,
\be
S_{PU} =\frac 12\xi \int \diff^4x\left(\Box\phi\right)^2=\frac 12\xi\int \delta\diff\phi\wedge*\delta\diff\phi\, .
\ee
Hence $\diff^4x\left(\Box\phi\right)^2=  \delta\diff\phi\wedge*\delta\diff\phi = *\diff*\diff\phi\wedge \diff*\diff\phi$.
Here $\de = -*\diff*$ is the co-differential (on a 4-dimensional Lorentz manifold) and $*\om$ denotes the Hodge dual of the
p-form $\om$ defined by
$$ (*\om)_{i_1\cdots i_{n-p}} = \frac{1}{p!}\eta^{i_1\cdots i_n}\om_{i_{n-p}\cdots i_n} \,, \quad \mbox{ here } \eta 
  ~\mbox{ is the volume form, } ~\eta =\sqrt{-g}\diff x^1\wedge \cdots \wedge \diff x^n \,.$$
Since $\de\de=0$, the PU action is invariant under the transformation  $\diff\phi \ra \diff\phi +\de\Si$ for 
some 2-form $\Si$
which has the property that $\de\Si = \diff\theta$ for some scalar field $\theta$. The field $\theta$  obviously satisfies 
$\Box\theta =\de \diff\theta =\de\de\Si =0$. This is nothing else that the remaining gauge invariance or  generalized shift 
symmetry of the PU action expressed in the language of forms.

\section{Interactions}\label{s:inter}
As we have already mentioned, it has been shown that the free PU oscillator can be quantized with a positive spectrum for the Hamiltonian. However, it happens very often that a free 
higher derivative theory can be fine but it becomes unstable once we introduce interactions. Actually, this is the real problem with higher order derivative theories.  Let us investigate this issue for our theory.
Since $\phi$ is the St\"uckelberg field of a $U(1)$ gauge field, its natural interaction will be to charged fields by means of a gauge interaction. In the following we shall consider two explicit examples, namely a complex scalar field and a charged fermion.
 
\subsection{Charged scalars}
Let us consider the action of a complex scalar field $\chi$ whose interaction is mediated by our St\"uckelberg field.
We define the covariant derivative $D_\mu=\partial_\mu-i\dd_\mu\phi$ so that the action has a $U(1)$ 
gauge symmetry. Under a gauge transformation $\chi \ra e^{i\theta}\chi$ and $\phi \ra \phi +\theta$, the 
following action is invariant
\begin{eqnarray}
S&=&\int \diff^4x\Big[ D_\mu\chi (D^\mu\chi)^*-V(\chi \chi^*)\Big]\nonumber\\
&=&\int \diff^4x\Big[ \partial_\mu\chi \partial^\mu\chi^*+\chi\chi^*\partial_\mu\phi\partial^\mu\phi-i\partial_\mu\phi\big(\chi\partial^\mu\chi^*-\chi^*\partial^\mu\chi\big)-V(\chi\chi^*)\Big]. \label{e:Schi}
\end{eqnarray}
Here $V$ is a potential which depends only on the modulus of $\chi$.
It is interesting to note that the scalar gauge field $\phi$ is automatically dynamical because it has derivative couplings to the charged field, unlike for the case of a gauge vector field. However, having a gauge symmetry, we can always remove it from action~(\ref{e:Schi}) by an appropriate gauge choice, so that it does not represent an actual physical degree of freedom of~(\ref{e:Schi}). Nevertheless, once we add 
the {\it free} PU action to the theory, $\phi$ can no longer be completely gauged away since the gauge symmetry is now reduced to gauge parameters satisfying $\Box\theta=0$. Thus, the full interacting Lagrangian is given by
\begin{eqnarray}
S=\int \diff^4x\left[ \partial_\mu\chi \partial^\mu\chi^*+\chi\chi^*\partial_\mu\phi\partial^\mu\phi-i\partial_\mu\phi\big(\chi\partial^\mu\chi^*-\chi^*\partial^\mu\chi\big)-V(\chi\chi^*)+\frac12 \xi(\Box\phi)^2\right]\,.
\end{eqnarray}
With this additional kinetic term for the gauge field $\phi$, it actually propagates two degrees of freedom, one of which can be removed using the gauge freedom so that $\phi$ does now propagate one physical degree of freedom.  Notice also that the addition of the PU term gives its corresponding propagator. 

It is also interesting to note that if the potential $V(\chi\chi^*)$ leads to symmetry breaking, we obtain a non-degenerate PU model from the non-vanishing vacuum expectation value (vev) of $\chi\chi^*$. For instance, if we consider the usual renormalizable quartic potential $V(\vert \chi\vert)=\mu^2\vert \chi\vert^2+\lambda\vert \chi\vert^4$ with $\mu^2<0$, then $\chi$ acquires a non-vanishing vev $\chi\chi^*\equiv v^2=-\mu^2/\lambda$ and the quadratic term for $\phi$ is given by 
\be
S^{(2)}_{\phi}=\int\diff^4x\left[\frac12\xi(\Box\phi)^2+v^2\partial_\mu\phi\partial^\mu\phi\right].
\ee
Thus, we can generate the term with two derivatives analogous to the PU action with $m_1^2=m_2^2=-2v^2/\xi$ from a spontaneous symmetry breaking of the charged field.

The equations of motion derived from the above action are\footnote{Here we use the fact that $D_\mu\chi^*=(D_\mu\chi)^*$ so that we will not distinguish between both. }
\begin{eqnarray}
\xi\Box^2\phi-\partial_\mu j^\mu&=&0, \\
D_\mu D^\mu\chi^*+\frac{\partial V}{\partial \chi}&=&0,\\
D_\mu D^\mu\chi+\frac{\partial V}{\partial \chi^*}&=&0 \,,
\end{eqnarray}
where we have introduced the current
\be
j^\mu= i\left(\chi^*D^\mu\chi - \chi D^\mu\chi^* \right) =  i \left(   \chi^* \partial^\mu \chi  - \chi \partial^\mu \chi^*\right) + 2 \chi \chi^* \partial^\mu \phi \,.
\ee
Notice that the first equation can be written as a conservation equation as follows:
\be
\partial_\mu\left(\xi\partial^\mu\Box\phi-j^\mu\right)=0
\ee
so that we have a conserved charge given by
\be
\mathcal{Q}=\int\diff^3x\left(\xi\Box\dot{\phi}-j^0\right).
\ee
Interestingly, since the $\chi$ sector has    a global $U(1)$ symmetry , the current $j^\mu$ is independently conserved on-shell, i.e., $\partial_\mu j^\mu=0$ so that we have the usual conservation of the complex field charge
\be
\mathcal{Q}_\chi=\int\diff^3x\;j^0.
\ee
This conservation law also implies the conservation of the current $j^\mu_\phi=\partial^\mu\Box\phi$ that gives rise to the conserved charge
\be
\mathcal{Q}_\phi=\int\diff^3x\;\Box\dot{\phi}
\ee
associated with the PU field. The crucial fact here is that the $\chi$-dependent term disappears from the equation of motion of $\phi$. However, the equation of motion for $\chi$ still depends on $\phi$. In other words, $\phi$ affects the dynamics of 
$\chi$ but is itself not affected by $\chi$. It satisfies a "free" equation even in the presence of the coupling to 
$\chi$. This surprising behavior will be useful for the quantization of the theory because we can use the 
procedure of the free field quantization to get rid of the ghost-mode and the coupling will not reintroduce
the ghost into the theory. 

To quantize the theory we write it in Hamiltonian form.
The conjugate momenta are given by
\begin{eqnarray}
\pi_\chi&=&\frac{\partial \mathcal{L}}{\partial\dot{\chi}}=D^0\chi^* , \\
\phi_1=\phi\,, \quad \Pi_1&=&\frac{\partial \mathcal{L}}{\partial\dot{\phi}}-\frac{\diff}{\diff t}\frac{\partial \mathcal{L}}{\partial\ddot{\phi}}= 2 \chi \chi^* \dot \phi - i \left( \chi \dot \chi^* - \dot \chi \chi^* \right) -  \xi \Box \dot \phi   =  j^0 -  \xi \Box \dot \phi,  \\
\phi_2=\dot\phi\,, \quad \Pi_2&=&\frac{\partial \mathcal{L}}{\partial\ddot{\phi}}=\xi\Box\phi .
\end{eqnarray}
The conjugate momentum of $\chi^*$ is of course $\pi_{\chi^*}=\pi_\chi^*=D^0\chi$. Thus, the Hamiltonian density is
\bea
\mathcal{H}&=&\dot{\chi}\pi_\chi+\dot{\chi}^*\pi_\chi^*+\dot{\phi}\Pi_1+\ddot{\phi}\Pi_2 - \mathcal{L} \nonumber \\
&=&D_0\chi D^0\chi^*-D_i\chi D^i\chi^*+V\left( \chi \chi^* \right) +\frac12\xi(\Box\phi)^2+\xi\left(\left( \nabla^2\phi \right) \Box\phi-\dot{\phi}\Box\dot{\phi}\right).
\eea
Writing this in terms of the momenta and the fields we obtain
\bea
\mathcal{H}(\chi,\phi_1,\phi_2,\pi_\chi,\Pi_1,\Pi_2) &=& \pi_\chi \pi_\chi^* + i \phi_2 \left( \chi \pi_\chi - \chi^* \pi_\chi^* \right) + \phi_2 \Pi_1 + \frac{\Pi_2^2}{2 \xi} +  \Pi_2  \nabla^2 \phi_1\nonumber \\
&&\hspace{-2.3cm}+ \nabla \chi \nabla \chi^* + \chi \chi^* \left( \nabla \phi_1 \right)^2 - i \nabla \phi_1 \left( \chi \nabla \chi^* - \chi^* \nabla \chi \right) + V\left( \chi \chi^* \right) .
\eea

\subsection{Charged fermions}
Here we shall briefly repeat the derivations of the previous section for  a coupling of $\phi$ to a Dirac field $\psi$. Again, by using the covariant derivative $D_\mu=\dd_\mu-i\dd_\mu\phi$,  the  action for a charged fermion becomes
\bea\label{e:dirac}
S&=& \int \diff^4x \left[ - \bar \psi \left( \gamma^\mu D_\mu + m \right) \psi   \right] \nonumber \\
&=& \int \diff^4x \left[ - \bar \psi \gamma^\mu \partial_\mu \psi + i \bar \psi \gamma^\mu\psi \partial_\mu \phi - m \bar \psi \psi  \right].
\eea
This action is invariant under a gauge transformation with $\psi \ra e^{i\theta}\psi$ and $\phi \ra \phi +\theta$. Again, having this symmetry at our disposal, the scalar  field $\phi$ can, in principle, be gauged away. However, when we identify it with the PU field, it actually carries two degrees of freedom, one of which will be physical. Thus, the full interacting Lagrangian is
 \be
S= \int \diff^4x \left[ - \bar \psi \gamma^\mu \partial_\mu \psi + i \bar \psi \gamma^\mu\psi \partial_\mu \phi - m \bar \psi \psi + \frac12\xi \left( \Box \phi \right)^2 \right].
 \ee
 This action leads to the following equations of motion
 \bea
 \xi \Box^2 \phi - \partial_\mu j^\mu &=&0, \\
 D_\mu \bar \psi \gamma^\mu - m \bar \psi &=&0, \\
 D_\mu  \gamma^\mu \psi + m  \psi &=&0,
 \eea
where we have introduced the usual $U(1)$ current $j^\mu = i \bar \psi \gamma^\mu \psi$, which is conserved, i.e., $\partial_\mu j^\mu=0$ so that we have the usual conserved fermionic charge
 \be
 {\mathcal Q}_\psi=i \int\diff^3x\bar\psi\gamma^0\psi . 
 \ee
Again, we also have the conserved current associated with the PU field ${\mathcal Q}_{\phi}$. Therefore, like for the case of a coupling to a charged scalar field, the PU field satisfies the "free" equation, i.e., its dynamics is not affected by the presence of the interaction with the fermionic field, however the PU field $\phi$ can, in principle, affects the dynamics of $\psi$. 
 
To end this section and for completeness, we shall compute the Hamiltonian. The corresponding conjugate momenta are given by
 \bea \label{pi_psi}
  \pi_\psi &=& \frac{\partial \mathcal{L}}{\partial \dot \psi}=  - \bar \psi \gamma^0 \quad \Rightarrow \ \bar \psi = \pi_\psi \gamma^0,\\
\phi_1=\phi\,, \quad \Pi_1&=& \frac{\partial \mathcal{L}}{\partial \dot \phi} - \frac{\diff}{\diff t} \frac{\partial \mathcal{L}}{\partial \ddot \phi} = i \bar \psi \gamma^0 \psi -  \xi \Box \dot \phi = j^0 -  \xi \Box \dot \phi, \\
\phi_2=\dot\phi\,, \quad \Pi_2 &=& \frac{\partial \mathcal{L}}{\partial \ddot \phi} =  \xi \Box \phi.
 \eea
As we notice from Eq.~(\ref{pi_psi}), the field $\bar \psi$ is proportional to the conjugate momentum $\pi_\psi$ and so we should not consider $\bar \psi$ as a field like $\psi$ (cf.~\cite{WeinbergQFT}).
The Hamiltonian density reads
\bea
 \mathcal{H} &=& \pi_\psi \dot \psi + \Pi_1 \dot \phi + \Pi_2 \ddot \phi - \mathcal{L}  \nonumber \\
 &=& \Pi_1 \phi_2 + \frac{\Pi_2^2}{2 \xi} +  \Pi_2 \nabla^2 \phi_1  + \pi_\psi \gamma^0 \left( m + \gamma^i \partial_i \right) \psi + i \pi_\psi \left( \phi_2 - \gamma^0 \gamma^i \partial_i \phi_1 \right) \psi.
\eea

In summary, in this Section we have shown explicitly that, by introducing interactions of the PU field to charged scalars or fermions  following a minimal coupling principle, the dynamics of the PU field is not modified. The reason for this is that, although the quadratic term giving the free propagator for $\phi$ only respects a residual gauge symmetry, the full theory still preserves global $U(1)$ symmetry that gives rise to  current conservation. Since it is the divergence of the conserved current that enters into the equation of motion of $\phi$ (which is guaranteed precisely by introducing it through a $U(1)$ covariant derivative), no effects from the charged particles on the PU field appear.

The above result that charged particles cannot excite the PU field can also been understood from standard electromagnetism results, where photons with polarization vector proportional to the 4-momentum $k_\mu$ cannot be generated out of conserved currents. At the quantum level, this is ensured by the Ward identities according to which the amplitude of any process involving an external longitudinal photon (in the 4-dimensional sense) vanishes.

It is interesting to note that the above procedure to gauge a global $U(1)$ symmetry is similar to making it local by introducing a {\it longitudinal} vector field. In fact, if we consider the covariant derivative $D_\mu=\partial_\mu-iA_\mu$, nothing here imposes that the vector field $A_\mu$ must be transverse and only once we choose the fully gauge invariant kinetic term $-F^2/4$ for $A_\mu$, the propagating vector boson becomes transverse. This is the natural choice if we want the gauge boson to carry a pure massless spin-1 representation of the Lorentz group and also if we want to keep the full $U(1)$ gauge invariance. However, other possibilities could be considered. One can, for instance, choose the kinetic term $(\partial_\mu A^\mu)^2/2$ for the vector field so that only its temporal component propagates. The price to pay is that only a residual gauge symmetry remains in the sector of the gauge boson. However, this is not too problematic in principle since  the global $U(1)$ symmetry is maintained ($A_\mu$ does not change under a global transformation) and charge conservation is not affected. Notice also that, in addition to the coupling of the charged fields to $\phi$, we can also couple them to the transverse gauge field $A_\mu$,
in principle with a different coupling constant, i.e., they can be differently charged under the {\it two different} $U(1)$ fields.

\section{Discussion}\label{s:disc}

In the previous sections we have introduced interactions for the degenerate PU field by following a symmetry principle, according to which all the interactions respect  $U(1)$ gauge symmetry and $\phi$ is identified with the gauge field that allows to render the gauge symmetry local. It is precisely this way of coupling the PU field that facilitates to consistently quantize the theory. We have shown that the interactions of the PU field with charged scalars or fermions do not modify the equation of motion for $\phi$, although $\phi$ itself can affect the dynamics of the charged fields. Thus, as we shall show below, one can quantize the free field and use the gauge symmetry to remove the ghost-like mode. The interactions will then not re-introduce it. After quantizing the field, we shall discuss its cosmological relevance and show how it can give rise to an effective cosmological constant.

\subsection{Quantization and Stability of the PU field}
A method to quantize the PU oscillator has been proposed in \cite{Bender:2007wu}. This method relies on the fact that even if a Hamiltonian is not hermitian, it leads to a unitary quantum theory if it exhibits an unbroken $\mathcal{PT}$-symmetry.  In the original approach, the quantization procedure was developed for an isolated PU oscillator. Of course, being isolated, the ghostly degree of freedom is harmless \cite{Smilga:2008pr}. One should actually check if couplings to other oscillators (or other dynamical systems) can be consistently added in such a way that the theory remains stable. In \cite{Bender:2007wb}, the coupling of hermitian and non-hermitian Hamiltonians was explored and it was shown that if the coupling constant is large enough, the energies can become complex. In some other works \cite{Smilga:2004cy}, higher derivative supersymmetric theories are studied and it is shown that some special types of interactions do not spoil the unitarity of the theories.  On the other hand, while the quantization method yields a system with positive energies, the classical Hamiltonian remains unbounded so that one needs to establish how the classical limit should be taken. Here we shall not make use of this quantization approach, but we shall take advantage of the existing residual gauge symmetry to get rid of the undesired degree of freedom.  Moreover, we shall discuss how the quantization is consistent in view of the results of the previous Section.

The Hamiltonian for the free PU field in terms of the conjugate momenta,
 \bea
 \Pi_1&=& \frac{\partial \mathcal{L}}{\partial \dot \phi} - \frac{\diff}{ \diff t} \frac{\partial \mathcal{L}}{\partial \ddot \phi} =  -  \xi \Box \dot \phi ,\\
 \phi_2=\dot\phi\,, \quad\Pi_2 &=& \frac{\partial \mathcal{L}}{\partial \ddot \phi} =  \xi \Box \phi,
 \eea
 is given by
\bea
\mathcal{H} _0=\Pi_1 \phi_2 + \frac{\Pi_2^2}{2\xi} + \Pi_2 \nabla^2 \phi \,,
\eea
where the expected Ostrogradski instability associated with $\Pi_1$ is represented by the unbounded first  term. This is not fatal by itself, as we have discussed above, but it can, in general, lead to instabilities when we couple it to another field. At the classical level, this instability can show up by the 
excitation of arbitrarily many modes coupling to $\phi$ at the cost of lowering $\mathcal{H} _0$ indefinitely.
As we have seen for the PU oscillator, we do not have any tachyonic instability, all the modes have real propagation speeds. Only for the degenerate case (that corresponds to our St\"uckelberg field) we have a mode that grows linearly with time, associated with having a double pole. This represents a much milder {\it instability} than those associated with tachyons. 

On the other hand, we notice that this Hamiltonian does not have the gauge symmetry of the theory. However, it will obviously lead to a set of Hamilton equations that do satisfy the symmetry. This should not be surprising since the Hamiltonian, in general,  does not preserve  symmetries. In particular, it does not respect Lorentz symmetry, although the corresponding theory does. It is the set of physical observables that must respect the symmetries of the theory. For instance, one can always perform a canonical transformation that will change the form of the Hamiltonian, but will leave the equations of the dynamical system invariant.  Moreover, in the case of a field theory, it is possible to add a 3-divergence or a total time derivative to the Hamiltonian density without modifying the quantum theory \cite{WeinbergQFT}.  In terms of the canonical variables, the gauge symmetry of the action reads $\Pi_{1,2}\rightarrow\Pi_{1,2}$, $\phi_1\rightarrow\phi_1+\theta$, $\phi_2\rightarrow\phi_2+\dot{\theta}$ with $\theta$ an arbitrary harmonic function. In principle, one could try to construct a gauge-invariant Hamiltonian by using the aforementioned allowed modifications, although we shall not pursue this approach here. Instead, we shall use the freedom given by the gauge symmetry to select a set of gauge-related modes for which the energy is positive.

For our equations of motion, the field can be expanded in Fourier modes as follows:
\be
\phi=\int\frac{\diff^3k}{(2\pi)^{3/2}}\frac{1}{(2k)^{3/2}}\left[(a_k+ib_kkt)e^{i(\vec{k}\cdot\vec{x}-kt)}+(a_k^*-ib_k^* kt)e^{-i(\vec{k}\cdot\vec{x}-kt)}\right].
\label{phidecomposition}
\ee
After quantization, $a_k$, $b_k$  and $a_k^*$, $b_k^*$ are promoted to operators $\aop_k$, $\bop_k$  and $\aop_k^\dag$, $\bop_k^\dag$ respectively. Notice that $a_k$ is the pure gauge mode since it is modified by  residual gauge transformations. In fact,  gauge-invariant quantities (that will be determined by $\Box\phi$) only depend on $b_k$ and $b_k^*$.

Using the above field decomposition, the Hamiltonian can be expressed as
\be
H=\int \diff^3x:\mathcal{H}_0:=\xi\int \diff^3k k\left[\bop_k^\dag \bop_k-\frac{1}{2}\left(\aop_k^\dag \bop_k+\bop_k^\dag \aop_k \right)\right],
\ee
where the double dots $:\ :$ denote the normal ordering operator. As anticipated, the Hamiltonian is not gauge invariant which is reflected by its dependence on the gauge mode $\aop_k$. Since we are free to perform a gauge transformation we could eliminate it from the physical spectrum by fixing some suitable gauge.   Below, we explicitly show how to proceed.  

Another approach is to impose an additional subsidiary condition \`a la Gupta-Bleuler\footnote{See for instance \cite{Itzykson}.} and to define a physical Hilbert space in which the expectation value of the gauge-dependent piece of the Hamiltonian vanishes. The easiest way to achieve this is to define the physical states as those which are annihilated by the gauge mode: $\aop_k\vert {\rm phys}\rangle=0$. In fact, this is too restrictive and can be relaxed, since imposing the more general condition $\left(\aop_k-i\alpha \bop_k\right)\vert{\rm phys}\rangle=0$ with $\alpha$ some arbitrary real number,  suffices to make the energy gauge invariant and positive for the physical Hilbert space. Moreover, if we only want to have positive energies, the parameter $\alpha$ of the subsidiary condition can also be complex, we just have to require that ${\rm Im}\; \alpha> -1$. This can be seen by computing the expectation value of the Hamiltonian in a physical state:
\be
\langle H\rangle_{\rm phys}=\xi(1+{\rm Im}\;\alpha)\int\diff^3kk\langle \bop_k^\dag \bop_k \rangle_{\rm phys}.
\ee
Notice that for the energy of the physical modes to be positive for $\alpha$ real we need $\xi>0$ and it has canonical energy if $\xi=1$.    The case with ${\rm Im} \;\alpha=-1$ in which the expectation value of the Hamiltonian vanishes is also interesting. Then, the gauge mode exactly cancels the energy of the physical mode, rendering this case equivalent to the original Gupta-Bleuler formalism for which the temporal and longitudinal modes of the electromagnetic potential cancel each other so that there is no contribution from them to the energy\footnote{In the standard Gupta-Bleuler approach to quantize electromagnetism, one requires the weak Lorenz condition so that the positive frequency part of field operator $\partial_\mu A^\mu$ annihilates the physical states, which guarantees that $\langle \partial_\mu A^\mu\rangle_{\rm phys}=0$. This is imposed in order to recover the classical Maxwell equations or, equivalently, so that only the transverse photons contribute to physical observables. However, as shown in \cite{Zwanziger} one could relax this condition and only require that physical states $\vert\psi\rangle$ are such that they have positive norm on physical observables $\langle {\mathcal O}\psi\vert{\mathcal O}\psi\rangle$, where ${\mathcal O}$ belongs to the observables algebra, defined as those operators that commute with the generator of the residual gauge symmetry.}.  The choice of the subsidiary condition with $\alpha$ real seems a natural choice because it fully eliminates the gauge-dependent piece of the Hamiltonian and it amounts to identifying a physical state as the entire equivalence class of states that differ only by a gauge mode. In terms of the PU field, the subsidiary condition with $\alpha$ real can be written as $\phi^{(+)}(t_0)\vert{\rm phys}\rangle=0$ where $\phi^{(+)}$ is the positive frequency part of the field operator and $t_0$ some given time. In other words, the physical states are those for which the expectation value of $\phi$ vanishes at some time, i.e., $\langle\phi\rangle_{\rm phys}=0$ for some $t=t_0$. 

For our bi-harmonic equation of motion, one can define the scalar product
\begin{eqnarray}
\Big(\phi_{k}, \phi_{k'}\Big)&=&i\int\diff^3x\Big[\Phi^{*T}_k\Pi_{k'}-\Pi^{*T}_k\Phi_{k'}\Big]\nonumber\\
&=&-i\xi \int\diff^3x\Big[ \left(\phi^*_k\Box\dot{\phi}_{k'}-\Box\dot{\phi^*}_k\phi_{k'}\right)- \left(\dot{\phi^*}_k\Box\phi_{k'}-\Box\phi^*_k\dot{\phi}_{k'}\right)  \Big]\label{scalarproduct}
\end{eqnarray}
where $T$ stands for the transpose and we have introduced the notation
\be
\Phi_k\equiv
\left( \begin{array}{c} \phi_k \\ \dot{\phi}_k
 \end{array} \right)\;\;\;\;\;\;\;\;\;\Pi_k\equiv
\xi\left( \begin{array}{c}- \Box{\dot{\phi}}_k \\ \Box\phi_k
 \end{array}\right).
\ee
The modes that we have used to decompose the field in (\ref{phidecomposition})  are not orthonormal with respect to this scalar product so that the corresponding operators do not satisfy the usual commutation relations, but they have the following commutation algebra:
\begin{eqnarray}
\Big[\aop_k,\aop_{k'}^\dag\Big]=\Big[\aop_k,\bop_{k'}^\dag\Big]=\Big[\bop_k,\aop_{k'}^\dag\Big]=-\frac{2}{\xi}\;\delta^{(3)}(\vec{k}-\vec{k}')
\end{eqnarray}
with all the other commutators vanishing. The modes that diagonalize the scalar product 
are\footnote{Notice the interesting appearance of the golden ratio $\tau=\frac{1+\sqrt{5}}{2}$ and its inverse $\tau^{-1}=-\frac{1-\sqrt{5}}{2}$.}:
\begin{eqnarray}
\phi_{1,k}&=&\frac{1}{(2\pi)^{3/2}}\frac{5^{-1/4}}{(2k)^{3/2}}\sqrt{\frac{2}{\xi}}\left(\frac{1-\sqrt{5}}{2}+ikt\right)e^{i(\vec{k}\cdot\vec{x}-kt)} \\
\phi_{2,k}&=&\frac{1}{(2\pi)^{3/2}}\frac{5^{-1/4}}{(2k)^{3/2}}\sqrt{\frac{2}{\xi}}\left(\frac{1+\sqrt{5}}{2}+ikt\right)e^{i(\vec{k}\cdot\vec{x}-kt)}
\end{eqnarray}
with eigenvalues $+1$ and $-1$ so that the modes $\phi_{2,k}$ have negative norm and this leads to a Hilbert space with indefinite metric. The PU field operator can be expanded as
\be
\phi=\int \diff^3k\sum_{\lambda=1,2}\Big(\aop_{\lambda, k}\phi_{\lambda,k}+\aop^\dag_{\lambda,k}\phi^*_{\lambda,k}\Big),
\ee
 where now  the modes are orthonormal with respect to the scalar product (\ref{scalarproduct}) and the annihilation and creation operators satisfy the commutation relations
\be
\Big[\aop_{\lambda,k},\aop^\dag_{\lambda', k'}\Big]=\eta_{\lambda,\lambda'}\delta^{(3)}(\vec{k}-\vec{k}')
\ee
with $\eta_{\lambda\lambda'}=$diag$(1,-1)$. Here we see how the negative norm modes appear in the commutation relations and lead to the indefinite metric for the corresponding Hilbert space. The relation of these operators with $\aop_k$ and $\bop_k$ is given by:
\begin{eqnarray}
\aop_k&=&\frac{5^{-1/4}}{\sqrt{2\xi}}\Big[\aop_{1,k}+\aop_{2,k}-\sqrt{5}(\aop_{1,k}-\aop_{2,k})\Big],\\
\bop_k&=&\sqrt{\frac 2\xi}5^{-1/4}\Big(\aop_{1,k}+\aop_{2,k}\Big).
\end{eqnarray}
Notice that the physical gauge-invariant mode $\bop_k$ is now given by $(\aop_{1,k}+\aop_{2,k})$, i.e., $\Box\phi$ is given in terms of such a combination. The hamiltonian in terms of these operators reads:
\be
H=\int\diff^3k\frac{k}{\sqrt{5}}\left[\left(1+\sqrt{5}\right)\aop^\dag_{1,k}\aop_{1,k}+\left(1-\sqrt{5}\right)\aop^\dag_{2,k}\aop_{2,k}+\left(\aop_{1,k}^\dag \aop_{2,k}+\aop_{2,k}^\dag \aop_{1,k}\right)\right]
\ee
Now we can impose the subsidiary condition $(\aop_k -i\alpha \bop_k)\vert{\rm phys}\rangle=0$ to obtain the physical Hilbert space. If we translate such a condition to the new operators it reads 
\be
\left[(1-\sqrt{5}-2i\alpha)\aop_{1,k}+(1+\sqrt{5}-2i\alpha)\aop_{2,k}\right]\Big\vert{\rm phys}\Big\rangle=0
\ee
and the expectation value of the Hamiltonian in a physical state is again gauge-independent and positive definite. Another way of quantizing the theory that takes advantage of the expansion in orthonormal modes is to fix the gauge such that we eliminate the negative norm mode and, then,  quantizing the positive norm mode alone, in analogy to the quantization in the Coulomb gauge for standard QED. To do this, one has to fix the   gauge $a_{2,k}=0$ a priori. This corresponds to choosing the gauge   mode $a_k=-\tau^{-1} b_k$ and $a_k^*=-\tau^{-1} b_k^*$, with $\tau$ the golden ratio, for the classical amplitudes. Then, the PU field operator is expanded in terms of the remaining positive modes, i.e.:
\be
\phi=\int \diff^3k\Big[\aop_{1,k}\phi_{1,k}+\aop_{1,k}^\dag\phi^*_{1,k}\Big].
\ee
Thus, these modes are orthogonal with norm $+1$ so that the annihilation and creation operators satisfy the usual commutation relations $[\aop_{1,k}, \;\aop_{1,k'}^\dag]=\delta^{(3)}(\vec{k}-\vec{k}')$. We thus avoid having to work with negative norm states and negative energies from the beginning without having to impose a subsidiary condition. The disadvantage of this quantization procedure will be the lost of explicit gauge-invariance. In the quantization \`a la Gupta-Bleuler previously mentioned we work with all the modes. The expectation values of the physical observables are explicitly gauge-independent because they will only depend on $\bop_k$ and $\bop_k^\dag$, which are the physical modes that cannot be removed by means of a gauge transformation.

The natural concern arising with the discussed procedure to identify the physical states is whether the introduction of interactions will spoil it. However, as we have shown in the previous section, if interactions are introduced following a symmetry principle, this will not be the case because, while the PU field will affect the sector of charged particles, its own equation of motion remains unaffected. The reason for this is that introducing the couplings as $U(1)$ gauge interactions leads to the presence of the associated Noether current, whose divergence is precisely the new term appearing in the equation of motion of the PU field. Thus, the conservation of this current also implies that  the equation of the PU field is not modified. This is the crucial point to guarantee that the discussed procedure to identify the physical states remains unaffected when introducing interactions because the field decomposition given in (\ref{phidecomposition}) is also valid (and exact) in the presence of interactions with charged particles.  The equivalent of this statement in standard QED comes from the fact that, due to current conservation, the divergence of the vector potential satisfies a decoupled free wave equation so that the selection of the Hilbert space is dynamically stable.

\subsection{Cosmological relevance}
Let us finally show how the PU field can actually play the role of dark energy. For that, we  first note that $\Box\phi$ satisfies the usual equation for a massless scalar field which, for a homogeneous field in a FLRW universe with metric  $ds^2=dt^2-a(t)^2d\vec{x}^2$, reads
\be
\left[\frac{\diff^2}{\diff t^2}+3H\frac{\diff}{\diff t}\right]\Box\phi=0.
\ee
The solution of this equation is given by
\be
 \Box\phi(t)=C_1+C_2\int\frac{\diff t}{a^3} 
 \ee
 where $C_{1,2}$ are integration constants. Since the $C_2$-mode decays throughout the expansion of the universe, only the constant mode $C_1$ is relevant at late time. This is simply the well known  result that a massless\footnote{More precisely, a scalar field with a mass much smaller than the Hubble expansion rate.} scalar field is frozen on super-Hubble scales. If we now compute the energy-momentum tensor for the PU field we obtain
\be
T_{\mu \nu}   =\xi g_{\mu \nu} \left( \frac{(\Box\phi) ^2}{2} + \dd_\lambda \phi \dd^\lambda \Box\phi  \right) - 2\xi\dd_{(\mu} \phi  \dd_{\nu)} \Box\phi  .
\ee
Thus, it becomes apparent why the PU field can drive an accelerated expansion. At late times, $\Box\phi$ is constant so that we obtain that the above energy-momentum tensor is simply
\be
T_{\mu\nu}=\frac\xi2\left(\Box\phi\right)^2g_{\mu\nu}\, .
\ee
This is the form of the energy-momentum tensor of a cosmological constant with the value $\La=4\pi G\xi(\Box\phi)^2$. Notice that  we must have $\xi>0$ for the effective cosmological constant to be positive. This is also what one would expect since, if $\Box\phi$ is constant, it contributes as a cosmological constant in the action. This result is in agreement with Ref.~\cite{Anisimov:2005ne} where the cosmology of a theory whose action is a general function of the D'Alembertian of a scalar field is explored and it is found that de Sitter is an attractor for this type of theories. This is also expected from the results of 
Refs.~\cite{Jimenez:2009dt}, where the symmetry breaking term $(\nabla_\mu A^\mu)^2$ gives rise to the appearance of an effective cosmological constant on super-Hubble scales. This relates to our result here because our PU field plays the role of the additional degree of freedom introduced by the symmetry breaking term there. 

Since the field $\Box\phi$ behaves like a massless scalar field, the primordial power spectrum generated from its quantum fluctuations during a de Sitter inflationary phase will be the usual scale invariant one given by\footnote{The  power of $H_I^4$ appearing here as opposed to the power $H_I^2$ found for the usual scalar field can be understood from dimensional arguments, but the underlying reason is that $\phi$ satisfies a fourth order equation so that it has a different normalization. Thus, although $\Box\phi$ satisfies the same equation as a massless scalar field, the power spectrum amplitude is different.} $P_{\Box\phi} \simeq\ H_I^4$ where $H_I$ is the constant Hubble parameter during inflation. Of course, in a more realistic quasi de Sitter inflationary phase, we expect a slightly tilted power spectrum with spectral index proportional to the slow roll parameters. However, the de Sitter expression will suffice for our purpose. 

Interestingly, the scale of the effective cosmological constant is set by the scale of inflation $M_p^2\Lambda\simeq(\Box\phi)^2\simeq H_I^4\simeq (M_I^2/M_p)^4$, with $M_I$ and $M_p$ the inflationary and Planck scales respectively and we have used the Friedmann equation to relate $H_I$ with $M_I$. Since we know that $\Lambda\simeq H_0^2$ where $H_0$ denotes the Hubble parameter today, we obtain 
 $M_I^4\simeq H_0M_p^3$. If we use the corresponding values for $H_0$ and $M_p$ we find 
 $M_I\simeq 1$TeV, i.e., the electroweak scale. This is the result that was also found in Ref.~\cite{Jimenez:2009dt} where the role of effective cosmological constant is played by $(\nabla_\mu A^\mu)^2$. Also, in \cite{Ringeval:2010hf} it is shown that quantum fluctuations of a scalar field during inflation can produce dark energy provided its mass is smaller than the Hubble expansion rate today. 
 
 This is a general feature of this type of models: If we have an action without any dimensionfull parameter that effectively gives a massless scalar field  and whose energy density is constant on super-Hubble scales, then the value of the effective cosmological constant that is generated during an inflationary phase taking place at the electroweak scale  coincides with the observed value. However, the scalar field must arise from some non-standard mechanism, since the energy density of a standard scalar field is diluted by the expansion of the universe. In the present case, the effective scalar field is the physical degree of freedom remaining in the degenerate PU model that is associated with the d'Alembertian of the field, while in \cite{Jimenez:2009dt} its role is played by the divergence of the vector field.
\newpage
\section{Conclusions and outlook}\label{s:con}
In this work we have shown that the degenerate Pais-Uhlenbeck field arises naturally as the St\"uckelberg field that restores the full $U(1)$ gauge symmetry in the action for a vector field including the usual Maxwell term plus a term $\propto (\dd_\mu A^\mu)^2$. We have seen that, in a given gauge, the St\"uckelberg field decouples from the vector field and satisfies the equation of motion of the degenerate PU field. Moreover, it determines the value of the longitudinal mode of the vector field so that it can be identified with it. After fixing to the decoupled gauge, we are left with the residual gauge symmetry analogous to the one remaining after imposing the Lorenz condition in standard electromagnetism. For the scalar field, this residual symmetry is an invariance under the addition of an arbitrary harmonic function so that it is a generalized version of the shift or galilean symmetries. 

 Although we have a fourth order theory, it propagates only one physical degree of freedom, since  the second degree of freedom can be removed by the residual symmetry. Also the ghost-like mode that one expects from a fourth order theory can be removed from the physical spectrum by an appropriate gauge choice so that the free Hamiltonian only contains the gauge-invariant and positive energy of the physical degree of freedom. This can be achieved easily in the case of the free field, however,  the introduction of interactions could spoil the mechanism by exciting the ghost-like mode. We have shown that if the couplings are introduced in a manner that the PU field acts as the gauge boson associated with a $U(1)$ symmetry, then the excising mechanism is not spoilt, i.e., the selection of the physical Hilbert space is dynamically stable. 
 
 We have explicitly worked out two particular cases:  coupling to a complex scalar field and  to a Dirac fermion. We have seen that in both cases, the equation of motion for the PU field remains unaffected thanks to the current conservation granted by the global $U(1)$ symmetry. This is crucial for the stability of the theory because guarantees that the ghost will not be reintroduced in the theory by the interactions   and the quantization procedure remains consistent even in the presence of interactions.  

Even though particles that are charged under such a $U(1)$ group are affected by the presence of the PU field, they cannot generate the PU field.  One way in which it could be produced is from quantum fluctuations during the inflationary era in the early universe. Once the quantum fluctuations are amplified, the super-Hubble modes contribute as an effective cosmological constant to the energy-momentum tensor whose scale is determined by the scale of inflation. The homogeneous evolution is exactly the same as in standard $\Lambda$CDM. However, as the cosmological term is truly the dynamical PU field, it can be perturbed and the evolution of the cosmological perturbations including it will differ from
standard $\Lambda$CDM. This provides us with a mechanism that can help to discriminate this model from a pure cosmological constant.

\section*{Acknowledgement}
We would like to thank Antonio L. Maroto for very useful discussions and comments. We also thank Andrei Smilga for useful remarks. This work is supported by the Swiss National Science Foundation. JBJ is supported by the Ministerio de Educaci\'on under the postdoctoral contract EX2009-0305 and the Spanish MICINNs Consolider-Ingenio 2010 Programme under grant MultiDark CSD2009-00064 and project number FIS2011-23000.


\begin{thebibliography}{99}

\bibitem{Komatsu}
E. Komatsu et~al.,  Astrophys. J. Suppl., 192, 18,  (2011) [arXiv:1001.4538];\\
D.~Larson et~al.,
 Astrophys. J. Suppl. {\bf 192}, 16 (2011), [arXiv:1001.4635].

\bibitem{nobel11}
S. Perlmutter et~al., Nature, \textbf{391}, 51 (1998);\\
S. Perlmutter et~al., Astrophys. J. \textbf{517}, 565 (1999);\\
A. G. Riess, A. V. Filippenko, P. Challis et al., Astron. J. \textbf{16}, 1009 (1998);\\
B. P. Schmidt, . et al., Astrophys. J. \textbf{507}, 46 (1998).

\bibitem{Martin:2012bt}
  J.~Martin,
  Comptes Rendus Physique {\bf 13}, 566 (2012)
  [arXiv:1205.3365 [astro-ph.CO]].

\bibitem{deRR}
R.~Durrer and R.~Maartens,
 Gen. Rel. Grav. {\bf 40}, 301 (2008), [arXiv:0711.0077],\\
DOI: 10.1007/s10714-007-0549-5.\\
  E.~J.~Copeland, M.~Sami and S.~Tsujikawa,
  Int.\ J.\ Mod.\ Phys.\ D {\bf 15} (2006) 1753
  [hep-th/0603057].\\
  S.~'i.~Nojiri and S.~D.~Odintsov,
  eConf C {\bf 0602061} (2006) 06
   [Int.\ J.\ Geom.\ Meth.\ Mod.\ Phys.\  {\bf 4} (2007) 115]
  [hep-th/0601213].

\bibitem{f(R)}
  T.~P.~Sotiriou and V.~Faraoni,
  Rev.\ Mod.\ Phys.\  {\bf 82} (2010) 451
  [arXiv:0805.1726 [gr-qc]].\\
  A.~De Felice and S.~Tsujikawa,
  Living Rev.\ Rel.\  {\bf 13} (2010) 3
  [arXiv:1002.4928 [gr-qc]].


\bibitem{galileon}
  A.~Nicolis, R.~Rattazzi and E.~Trincherini,
  Phys.\ Rev.\  D {\bf 79}, 064036 (2009)
  [arXiv:0811.2197 [hep-th]];\\
C. de Rham, L. Heisenberg,
Phys. Rev. {\bf D84}, 043503 (2011) [arXiv:1106.3312],\\
DOI: 	10.1103/PhysRevD.84.043503;\\
C.~de Rham and A.~J.~Tolley,
  JCAP {\bf 1005}, 015 (2010)
  [arXiv:1003.5917 [hep-th]].\\
  C.~Burrage, C.~de Rham and L.~Heisenberg,
  JCAP {\bf 1105} (2011) 025
  [arXiv:1104.0155 [hep-th]].
 
\bibitem{Ostro} M. Ostrogradski, Memoire Academie St. Petersbourg,
        Ser. VI {\bf 4}, 385 (1850).

\bibitem{Woody} 
  R.~P.~Woodard,
  Lect.\ Notes Phys.\  {\bf 720}, 403 (2007)
  [astro-ph/0601672].
  
 \bibitem{Chen:2012au}
  T.-j. Chen, M. Fasiello, E. A. Lim, A. J. Tolley,
  arXiv:1209.0583 [hep-th].

\bibitem{ArkaniHamed}
A. Adams et al., JHEP 0610:014 (2006) [arXiv:hep-th/0602178],\\
DOI: 	10.1088/1126-6708/2006/10/014

\bibitem{Jimenez:2009dt} 
 J.~Beltran Jimenez and A.~L.~Maroto,
  Mod.\ Phys.\ Lett.\ A {\bf 26}, 3025 (2011)
  [arXiv:1112.1106 [astro-ph.CO]].
  J.~Beltran Jimenez and A.~L.~Maroto,
  Phys.\ Lett.\ B {\bf 686}, 175 (2010)
  [arXiv:0903.4672 [astro-ph.CO]].
  J.~Beltran Jimenez and A.~L.~Maroto,
  JCAP {\bf 0903}, 016  (2009)
  [arXiv:0811.0566 [astro-ph]].
  J.~Beltran Jimenez and A.~L.~Maroto,
  Int.\ J.\ Mod.\ Phys.\ D {\bf 18} (2009) 2243
  [arXiv:0905.2589 [physics.gen-ph]].


   
\bibitem{Pais:1950za}
  A.~Pais and G.~E.~Uhlenbeck,
  Phys.\ Rev.\  {\bf 79},  145 (1950).

\bibitem{Bender:2007wu}
  C.~M.~Bender and P.~D.~Mannheim,
  Phys.\ Rev.\ Lett.\  {\bf 100}, 110402  (2008)
  [arXiv:0706.0207 [hep-th]].
  P.~D.~Mannheim,
  Found.\ Phys.\  {\bf 37} (2007) 532
  [hep-th/0608154].

\bibitem{Smilga:2008pr}
  A.~V.~Smilga,
  SIGMA {\bf 5} (2009) 017
  [arXiv:0808.0139 [quant-ph]].\\

\bibitem{Smilga:2004cy}
  A.~V.~Smilga,
  Nucl.\ Phys.\ B {\bf 706} (2005) 598
  [hep-th/0407231].
  D.~Robert and A.~V.~Smilga,
  J.\ Math.\ Phys.\  {\bf 49} (2008) 042104
  [math-ph/0611023].

\bibitem{Dirac}
P. A. M. Dirac, Can. J. Math., {\bf 2} (1950);\\
P. A. M. Dirac, {\em Lectures on Quantum Mechanics},  Courier Dover Publications (2001).

\bibitem{PUoscillator}
  C.~M.~Bender and P.~D.~Mannheim,
  arXiv:0804.4190 [hep-th].\\
  C.~M.~Bender and P.~D.~Mannheim,
  J.\ Phys.\  {\bf 41} (2008) 304018
  [arXiv:0807.2607 [hep-th]].





\bibitem{Stuck}
E. C. G. Stueckelberg, Helv. Phys. Acta {\bf 11}, 225 (1938); ibid. 225, ibid. 312.



\bibitem{WeinbergQFT}
S.~Weinberg,{\em The Quantum Theory of Fields}, Cambridge University Press (1995).



\bibitem{Bender:2007wb}
  C.~M.~Bender and H.~F.~Jones,
  J.\ Phys.\ A {\bf 41} (2008) 244006
  [arXiv:0709.3605 [hep-th]].




\bibitem{Itzykson} C. Itzykson and J.B. Zuber, {\it Quantum Field Theory},
McGraw-Hill (1980); N.N. Bogoliubov and D.V. Shirkov, {\it
Introduction to the theory of quantized fields}, Interscience
Publishers, Inc. (1959).

\bibitem{Zwanziger}
  D.~Zwanziger,
  Phys.\ Rev.\ D {\bf 14} (1976) 2570.

\bibitem{Anisimov:2005ne}
  A.~Anisimov, E.~Babichev and A.~Vikman,
  JCAP {\bf 0506}, 006  (2005)
  [astro-ph/0504560].
 
 
\bibitem{Ringeval:2010hf}
  C.~Ringeval, T.~Suyama, T.~Takahashi, M.~Yamaguchi and S.~Yokoyama,
  Phys.\ Rev.\ Lett.\  {\bf 105} (2010) 121301
  [arXiv:1006.0368 [astro-ph.CO]].


\end{thebibliography}
\end{document}